\documentclass[10pt]{iopart}

\usepackage{iopams}
\usepackage{graphicx}
\begin{document}

\title{Dynamic scaling and temperature effects in thin film roughening}

\author{T. A. de Assis}
\address{Instituto de F\'{\i}sica, Universidade Federal da Bahia,
   Campus Universit\'{a}rio da Federa\c c\~ao,
   Rua Bar\~{a}o de Jeremoabo s/n,
40170-115, Salvador, BA, Brazil}
\address{Instituto de F\'\i sica, Universidade Federal Fluminense, Avenida Litor\^anea s/n,
24210-340 Niter\'oi RJ, Brazil}
\ead{thiagoaa@ufba.br}

\author{F. D. A. Aar\~{a}o Reis}
\address{Instituto de F\'\i sica, Universidade Federal Fluminense, Avenida Litor\^anea s/n,
24210-340 Niter\'oi RJ, Brazil}
\ead{reis@if.uff.br}
\date{\today}

\begin{abstract}

The dynamic scaling of mesoscopically thick films (up to $10^{4}$ atomic layers) grown with
the Clarke-Vvedensky model is investigated numerically for broad ranges of values
of the diffusion-to-deposition ratio $R$ and lateral neighbor detachment probability $\epsilon$,
but with no barrier at step edges.
The global roughness scales with the film thickness
$t$ as $W \sim t^{\beta}/\left[R^{3/2}\left(\epsilon + a\right)\right]$, where
$\beta \approx 0.2$ is the growth exponent consistent with Villain-Lai-Das Sarma (VLDS)
scaling and $a=0.025$. This general dependence on $R$ and $\epsilon$ is inferred from
renormalization studies and shows a remarkable effect of the former but a small effect of
the latter, for $\epsilon\leq 0.1$. For $R\geq {10}^4$, very smooth surfaces are always produced.
The local roughness shows apparent anomalous scaling for very low temperatures
($R\leq {10}^2$), which is a consequence of large scaling corrections to asymptotic normal
scaling. The scaling variable $R^{3/2}\left( \epsilon + a\right)$ also represents
the temperature effects in the scaling of the correlation length and appears in the
dynamic scaling relation of the local roughness, which gives dynamic exponent $z\approx 3.3$
also consistent with the VLDS class.
\end{abstract}

\maketitle

\section{Introduction}
\label{intro}

Stochastic modeling of molecular-beam epitaxy (MBE) attracted much interest in the
last decades due to the importance of this technique to produce high quality
thin films for many applications \cite{ohring}. Those models adopt simple rules
for the aggregation, diffusion, and desorption processes, consequently
allowing the study of morphological properties of large samples \cite{barabasi,krug,etb}.
In the simplest
cases, they assume limited mobility (LM) of adatoms; some examples are the models
of Wolf and Villain \cite{wv} and of Das Sarma and Tamborenea \cite{dt}, in which
short-range surface diffusion and permanent aggregation take place immediatly after adsorption.
More realistic models consider thermally activated microscopic processes and
are frequently called colletive diffusion models. The most prominent example
is the Clarke-Vvedensky (CV) model \cite{cv}, in which the adatom diffusion
coefficients have Arrhenius forms, with energy barriers depending on the local
number of lattice neighbors. An important difference from LM models is that the
CV model obeys detailed balance conditions, thus it may also be a reliable description
of the film dynamics without deposition \cite{krug,siegert1992}.

In the basic formulation of the CV model,
the dynamics may be represented by temperature-like parameters
$R$ and $\epsilon$, respectively representing the diffusion-to-deposition ratio of
isolated atoms in terraces and the detachment probability at step edges \cite{etb,cv}.
For the description of specific MBE processes, at submonolayer or multilayer regime,
the models usually include additional
energy barriers for diffusion across step edges (upward and downward movements) and
additional adatom interactions;
a thorough review of homoepitaxy applications is presented in Ref.
\protect\cite{etb} and recent advance is discussed in
Refs. \protect\cite{clancy2011,einax,ferrando,Bommel}.
The formation of patterns, growth and coarsening of mounds were some of the features
that attracted much interest.

However, a small number of works analyzed the dynamic scaling of surface roughness
in CV-type models.
This analysis provides a set of scaling exponents connecting
the model to stochastic growth equations \cite{barabasi}, which helps
to distinguish the essential physico-chemical mechanisms of film growth.
Initial works on the basic CV model suggested temperature-dependent exponents and
anomalous scaling of surface roughness \cite{td,lanczycki,kotrla1996,meng}.
Subsequently, renormalization studies \cite{hasel2007,haselPRE2008,haselIJMPB2008}
suggested that it belongs to the class of the Villain-Lai-Das Sarma (VLDS)
growth equation \cite{villain,laidassarma}. In $2+1$ dimensions, VLDS scaling
was numerically confirmed for $\epsilon =0$ and a broad range of $R$
in Ref. \protect\cite{cdia} and for some values of $R$ and $\epsilon >0$ in
Ref. \protect\cite{lealJPCM}.

The aim of this paper is to perform a systematic investigation of dynamic
scaling in the basic CV model (without additional barriers across edges),
with particular attention on the role of the temperature-like parameters.
First, we study the scaling of global and local surface roughness and
confirm that it belongs to the VLDS class for a broad range of values of $R$ and $\epsilon$.
Deviations from this scaling are shown to appear only when the surfaces are very flat.
Second, we will show evidence of asymptotic normal scaling, but with an apparent
anomaly for short times and small $R$ similar to other VLDS models \cite{anomvlds}.
Finally, we will show that the roughness scales with $R$ in a form similar
to the irreversible aggregation model, with a weak dependence on the step detachment
rate $\epsilon$. Although this model without barriers at step edges is of limited
applicability to real solid films, these results may help the analysis of dynamic
scaling in extended versions of the CV model, particularly if crossover features
have to be analyzed.

The rest of this paper is organized as follows. In Sec. \ref{basic}, we present the model
and the related growth equation. In Sec. \ref{global}, we discuss
the dynamic scaling of the global surface roughness.
In Sec. \ref{local}, the scaling of local surface roughness is analyzed.
In Sec. \ref{conclusion}, we present our conclusions.

\section{Basic definitions and concepts}
\label{basic}

\subsection{Model and simulations}
\label{model}

The CV model is defined in a simple cubic lattice, with an initially flat substrate
at $z=0$. Deposition occurs with a flux of $F$ atoms per site per unit time,
in the $z$ direction towards the substrate. Each adatom occupies one lattice site, whose side
is taken as the unit length. We impose the solid-on-solid condition (i. e. overhangs
are not allowed), thus only adatoms at the top of each substrate column are mobile.

The hopping rate of an adatom with no lateral neighbor is
\begin{equation}
D_0=\nu_0\exp{\left( -E_s/k_BT\right)}
\label{defD0}
\end{equation}
where $\nu_0$ is a frequency, $E_s$ is an activation energy, and $T$ is the temperature.
The adatom step occurs in a randomly chosen substrate direction ($\pm x$, $\pm y$),
towards the top of a NN column.
If an adatom has $n$ lateral neighbors, its hopping rate is
\begin{equation}
D=D_0\epsilon^n \qquad , \qquad \epsilon \equiv \exp{\left( -E_b/k_BT\right)} ,
\label{defD}
\end{equation}
where $E_b$ is a bond energy.

An important parameter of the model is the diffusion-to-deposition ratio
\begin{equation}
R \equiv \frac{D_0}{F} = \frac{\nu_0}{F} \exp{\left( -E_s/k_BT\right)} .
\label{defR}
\end{equation}
It is usually interpreted as the number of steps of an adatom in a terrace before
it is buried by the next atomic layer. However, it is highly probable that an adatom
meets a lateral neighbor before being buried, which restricts that interpretation \cite{cdia}.
In the original CV model, $\nu_0=2k_BT/h$, where $h$ is the Planck's constant,
as predicted by transition state theory \cite{cv}. Some authors adopted that form
\cite{smilauer}, but it is more frequent that a constant value $\nu_0\sim {10}^{12}s^{-1}$
is considered in simulation and analytical works \cite{etb}. Here we will follow this
trend and consider a fixed ratio $\nu_0/F ={10}^{13}$.

Our simulations will be limited to deposition of ${10}^4$ monolayers. It is a typical
value for thin films, corresponding to thicknesses of order $2-3\mu m$, possibly more
for molecular materials. The simulation time $t$ will be expressed in number of deposited
layers. The substrate size is $L=1024$, which is large enough to avoid finite-size
effects in the chosen deposition time.
The values of $E_s$ and $E_b$ are determined by material properties.
Thus, since $R = \frac{\nu_0}{F}\epsilon^{E_s/E_b}$, the parameters $R$ and $\epsilon$
simulataneously vary with the temperature for a given material.
However, here we are interested in exploring a variety of physico-chemical conditions,
which include different values of activation energies. For this reason, $R$ and
$\epsilon$ will be taken as the independent parameters of the model.

We will perform simulations in the range $10\leq R\leq {10}^4$. Larger
values are expected in many MBE processes, but we will show that they produce very smooth
surfaces up to the maximal simulated thicknesses. We will also restrict our study
to $\epsilon\leq 0.1$, since larger values of this parameter would represent a solid
close to the melting point.

\subsection{Dynamic scaling and universality classes}

The main quantity to characterize the film surface is the local roughness
$w\left( r,t\right)$ in boxes of size $r$ at time $t$.
For calculating this quantity,
a square box of lateral size $r$ glides along the film surface and, at each position, the
root-mean-square (rms) height fluctuation of columns inside the box is calculated.
The average among all box positions and among different configurations
of the film at time $t$ is the local roughness.
The global roughness $W\left( t\right)$ is measured in the full system size $L$, i. e.
$W\left( t\right)=w\left( L,t\right)$.
In this work, very large substrates are considered, thus $L$ has negligible effect on $W$.

In systems with normal roughening (in opposition to anomalous roughening \cite{ramasco}),
the expected scaling of the local roughness in large substrates is
\begin{equation}
w\left( r,t\right) = r^{\alpha} f{\left( \frac{r}{t^{1/z}}\right)} ,
\label{fvlocal}
\end{equation}
where $\alpha$ and $z$ are the roughness and dynamic exponents, respectively, and
$f$ is a scaling function. For $x\equiv r/t^{1/z} \ll 1$ (small box sizes), $g(x)$ is constant;
for $x\gg 1$ (large box sizes), the local roughness converges to the global one,
$W\left( t\right)$, which scales as
\begin{equation}
W \sim t^{\beta} ,
\label{defbeta}
\end{equation}
where $\beta = \alpha /z$ is the growth exponent.

When growth is dominated by surface diffusion, it is expected to be
described by a fourth order stochastic equation in the hydrodynamic limit \cite{barabasi}:

\begin{equation}
{{\partial h(\vec{r},t)}\over{\partial t}} = \nu_4{\nabla}^4 h +
\lambda_{4} {\nabla}^2 {\left( \nabla h\right) }^2 + \eta (\vec{r},t) ,
\label{vlds}
\end{equation}
where $h(\vec{r},t)$ is the height at position $\vec{r}$ and time $t$ in a
$d$-dimensional substrate, $\nu_4$ and $\lambda_{4}$ are constants and $\eta$
is a Gaussian, nonconservative noise (the contribution of the average
external flux is omitted in Eq. \ref{vlds}).
The linear version ($\lambda_{4}=0$) is the Mullins-Herring (MH) equation \cite{mh},
while the nonlinear case is the VLDS equation \cite{villain,laidassarma}.

For the VLDS class in $2+1$ dimensions, the best estimates of scaling exponents are given by
the conserved restricted solid-on-solid models \cite{crsosreis}, and are very close to
one-loop renormalization values \cite{janssen}:
$\alpha \approx 2/3$, $z\approx 10/3$, and $\beta\approx 1/5$.

\section{Global roughness scaling}
\label{global}

\begin{figure}[!h]
\begin{center}
\includegraphics [width=8.5cm] {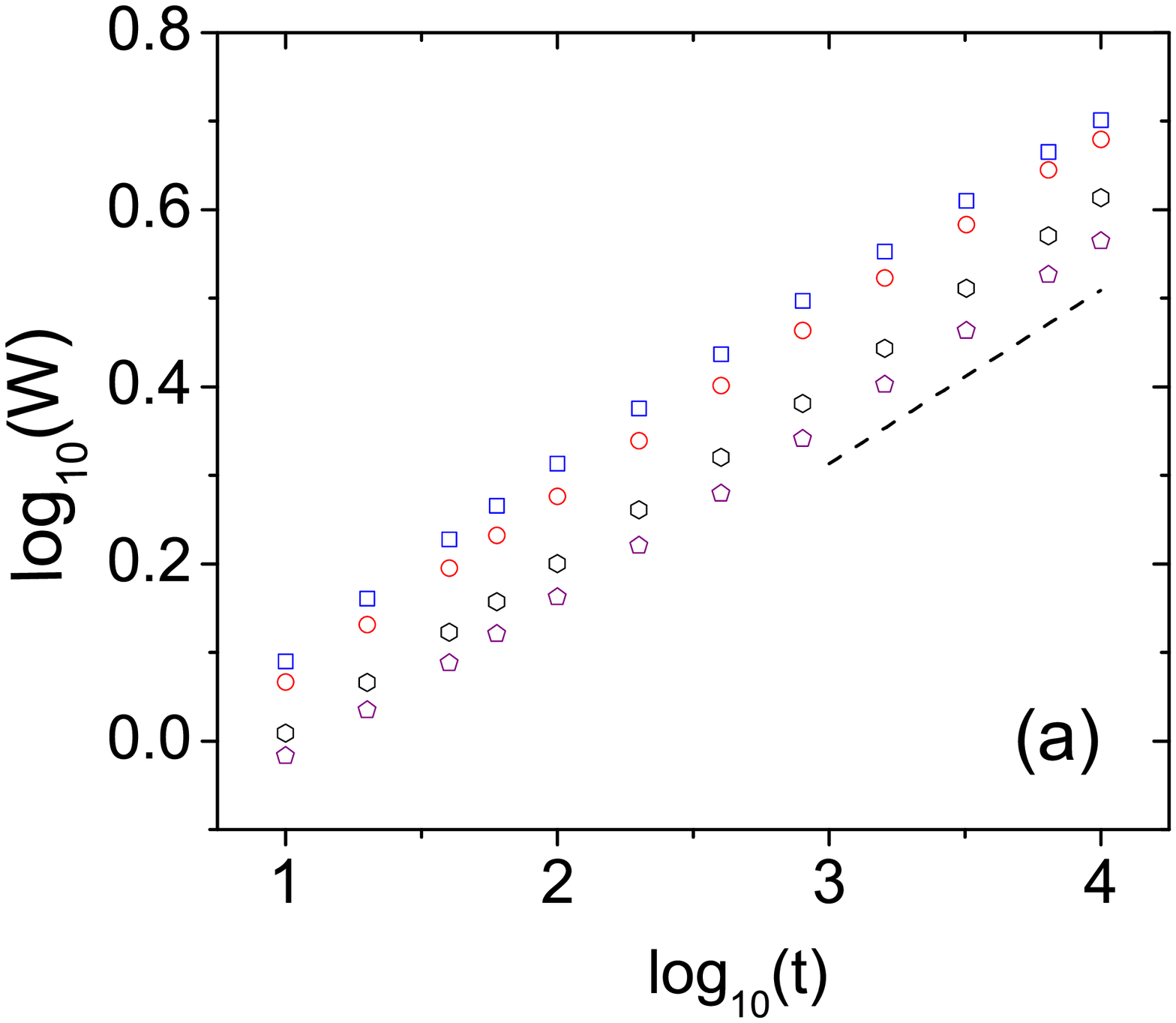}
\includegraphics [width=8.5cm] {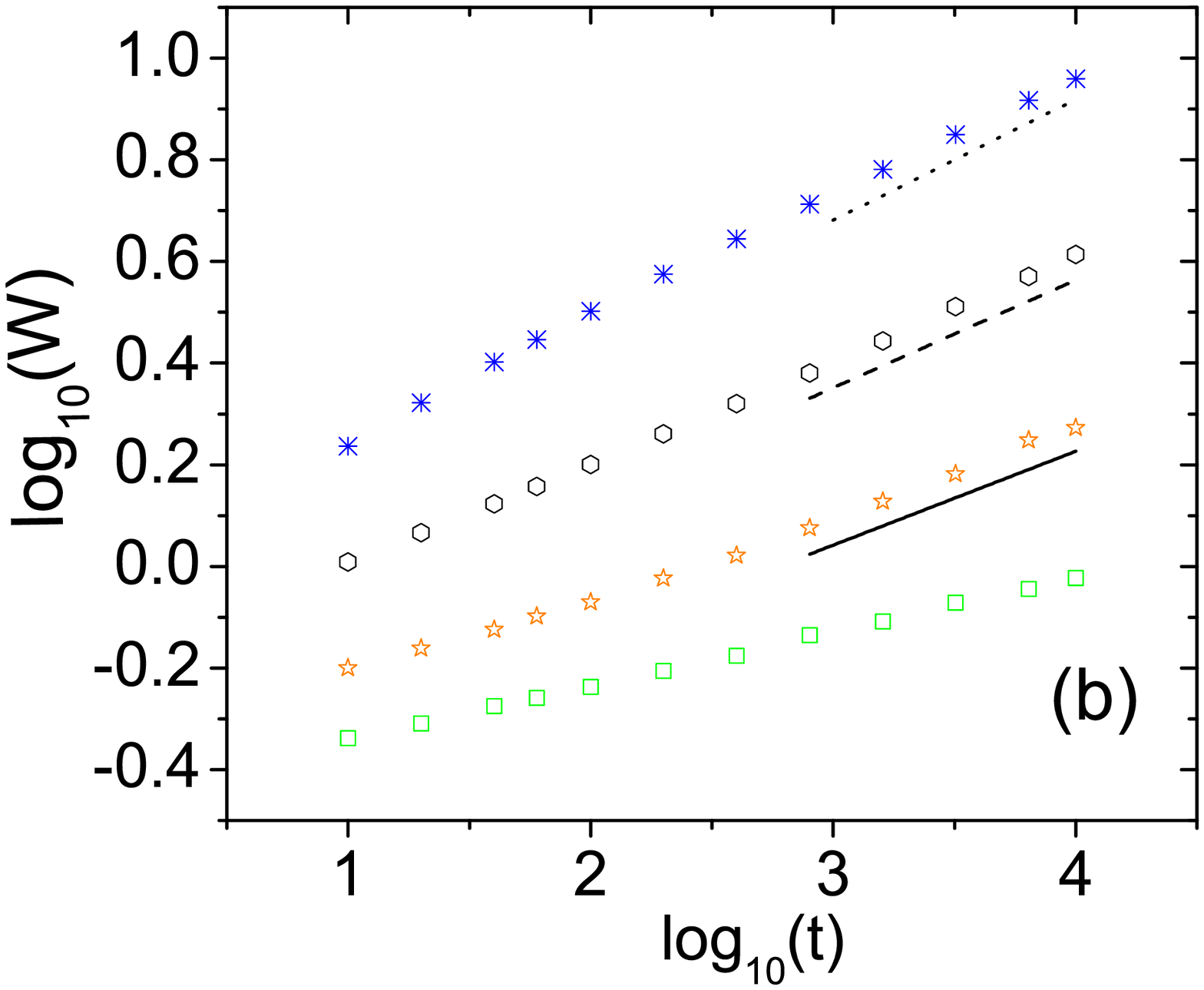}
\caption{(a) Global roughness as a function of time for $R={10}^2$, with $\epsilon =0$ (blue squares),
$0.01$ (red circles), $0.05$ (black hexagons), and $0.1$ (violet pentagons).
The dashed line has slope $0.20$.
(b) Global roughness as a function of time for $\epsilon=0.05$, with $R=10$ (blue asterisks),
${10}^2$ (black hexagons), ${10}^3$ (orange stars), and ${10}^4$ (green squares).
The dotted/dashed/full line has slope $0.22$/$0.20$/$0.18$.
}
\label{wglobal}
\end{center}
\end{figure}

In Fig. \ref{wglobal} (a),we show the roughness evolution for $R={10}^2$ and
several values of $\epsilon$. The time scaling gives $\beta\approx 0.20$.
As expected, the trend is that $W$ decreases as $\epsilon$ increases, since the
detachment from steps of atoms with one or two lateral bonds helps them to move
to positions with lower energy, forming more compact configurations.
However, the quantitative effect of $\epsilon$ on the roughness is small.

In Fig. \ref{wglobal} (b), we show the roughness evolution
for several values of $R$, with $\epsilon =0.05$.
In order to estimate the exponent $\beta$, fits of the data are done only for $W\geq 2$,
because smaller roughness corresponds to very smooth surfaces.
Those fits give $0.18\leq \beta\leq 0.22$, in good agreement with the VLDS exponent.
For $R={10}^4$, the roughness is very small at all times simulated here, thus
deviations appear.

As shown in Ref. \protect\cite{cdia}, for $\epsilon =0$, $W$ exceeds $2$ units only at
$t\sim {10}^5$; for $\epsilon >0$, this occurs for longer times.
Thus, for the values $R\geq {10}^4$ typical of most MBE processes, the film surfaces
grown with the basic CV model are very flat. The formation of patterns (e. g. mounds)
observed in many simulations is possible only with energy barriers at step edges.

Now we analyze the combined effects of parameters $R$ and $\epsilon$ on the
roughness scaling. In the case $\epsilon =0$, Ref. \protect\cite{cdia} showed that
\begin{equation}
W \sim \frac{t^{0.2}}{R^{0.3}}
\label{Wcdia}
\end{equation}
in the growth regime. This was derived from a Family-Vicsek relation \cite{fv}
that proposed the correlation length as $\xi\sim {\left( Rt\right)}^{1/z}$
[due to the subdiffusive propagation of correlations; $R\propto D$ from Eq. (\ref{defR})]
and the saturation roughness as $W_s\sim  R^{-1/2}$ (due to the formation
of plateaus of width $R^{1/z}$) \cite{cdia}.

Growth with $\epsilon >0$ allows the detachment of adatoms from steps. It helps
filling narrow surface valleys, which reduces the roughness. However, detachment may
occur at upward and downward edges of the plateaus, thus is has a small contribution
to their size. This explains why $\epsilon >0$ does not lead to drastic changes in
the roughness.
Consequently, the aggregation of free adatoms at the step edges is still the main
mechanism to determine the size of plateaus, and
the dependence of $W$ on $R$ is expected to be the same of the
model with irreversible step aggregation [Eq. (\ref{Wcdia})].

The renormalization study of the CV model helps to infer the dependence of the
roughness on the detachment probability. Ref. \protect\cite{haselPRE2008} derived the
coefficients of the corresponding Langevin equation (with VLDS form) as
a function of $D$ and $\gamma\equiv 1-\epsilon$. Those coefficients are
products of $D$ or $D\gamma$ by factors of the form ${\left( A+B\epsilon\right)}^c$,
where $A$ and $B$ are constants that depend on the coefficients of
regularization of step functions (which are sensitive to discrete model rules)
and $c$ is an integer.

\begin{figure}[!h]
\begin{center}
\includegraphics [width=8.5cm] {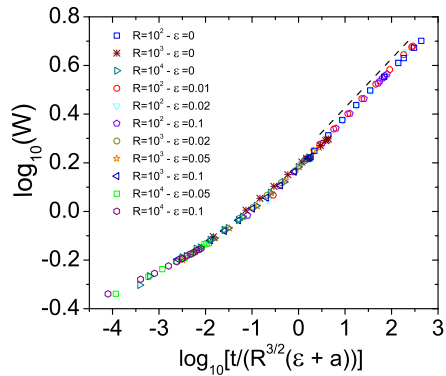}
\caption{Global roughness as a function of scaled time according to Eq. (\ref{WCV}),
with $a=0.025$. The dashed line has slope $0.2$.}
\label{Wscaling}
\end{center}
\end{figure}
Following this reasoning, we propose a scaling relation for the CV roughness as
\begin{equation}
W \sim \Im \left[ \frac{t}{R^{3/2}{\left( \epsilon + a\right)}^\theta} \right] ,
\label{WCV}
\end{equation}
where $\Im$ is a scaling function and $a$ and $\theta$ are constants.
We did not restrict the scaling to the power-law regime [Eqs. (\ref{defbeta})
or (\ref{Wcdia})] because this form may include the initial roughening (small $W$).

Fig. \ref{Wscaling} shows the roughness for several values of $R$
and $\epsilon$ with time scaled according to Eq. (\ref{WCV}).
The excellent data collapse
with $\theta = 1$ and $a=0.025$ confirms the proposed scaling relation.
For $\epsilon\lesssim a$, this relation clearly shows the weak dependence of $W$
on that parameter; for $\epsilon <{10}^{-3}$ ($E_b>7k_BT$), the effect of the detachment
rate is negligible.
The scaling of Eq. (\ref{WCV}) also includes the case $\epsilon =0$, in which
detailed balance fails.

\begin{figure}
\begin{center}
\includegraphics [width=5.5cm] {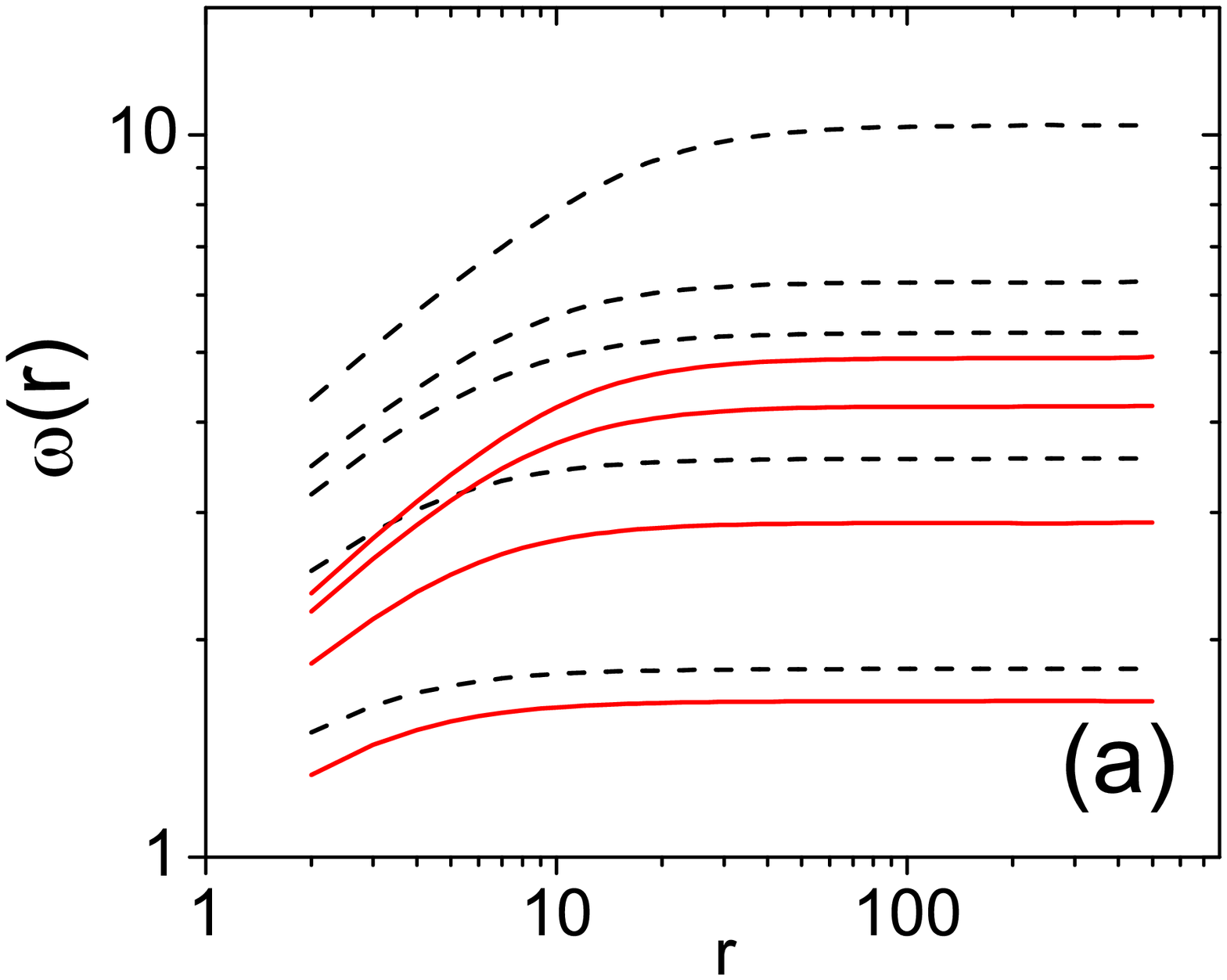}
\includegraphics [width=5.5cm] {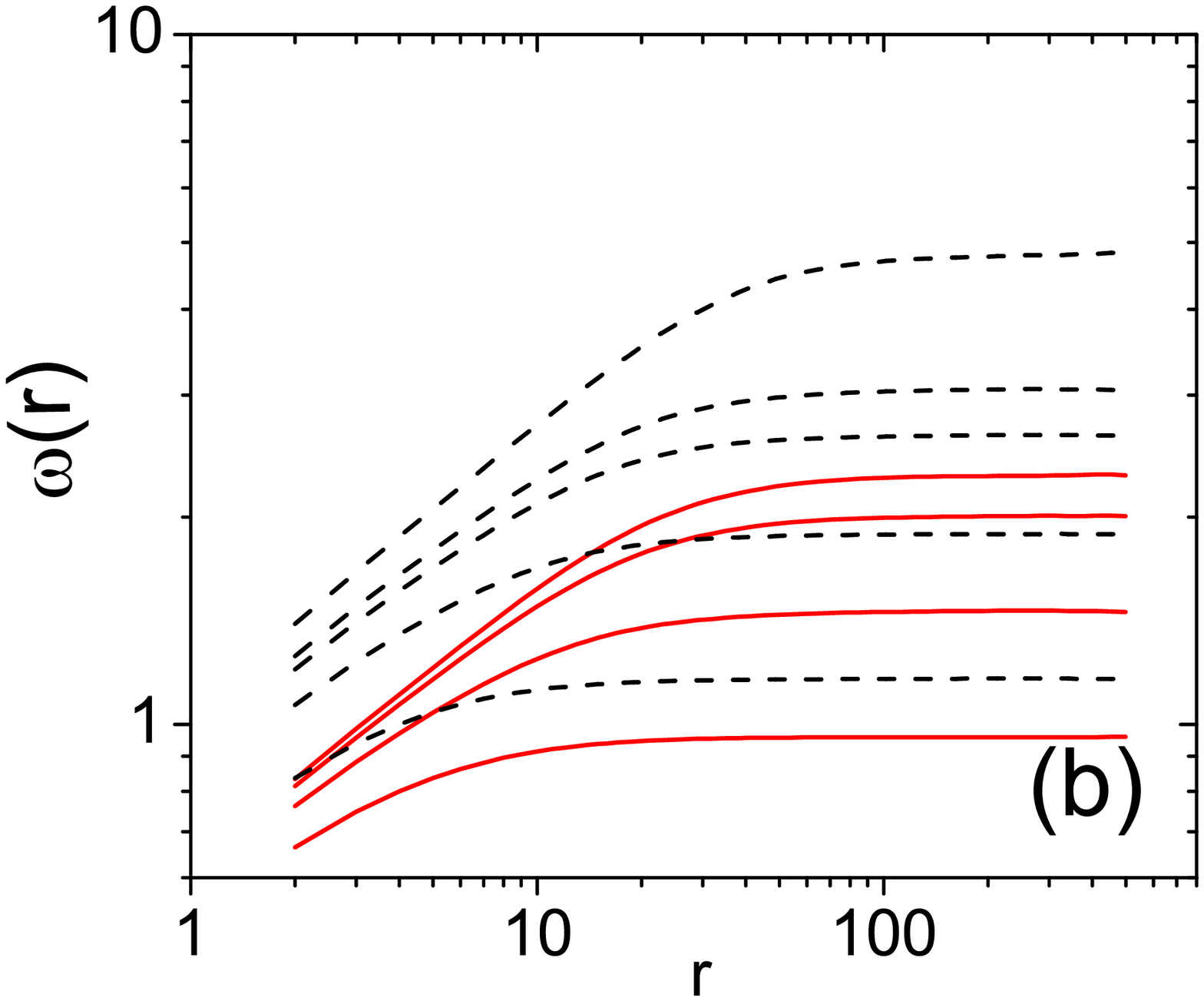}
\includegraphics [width=5.5cm] {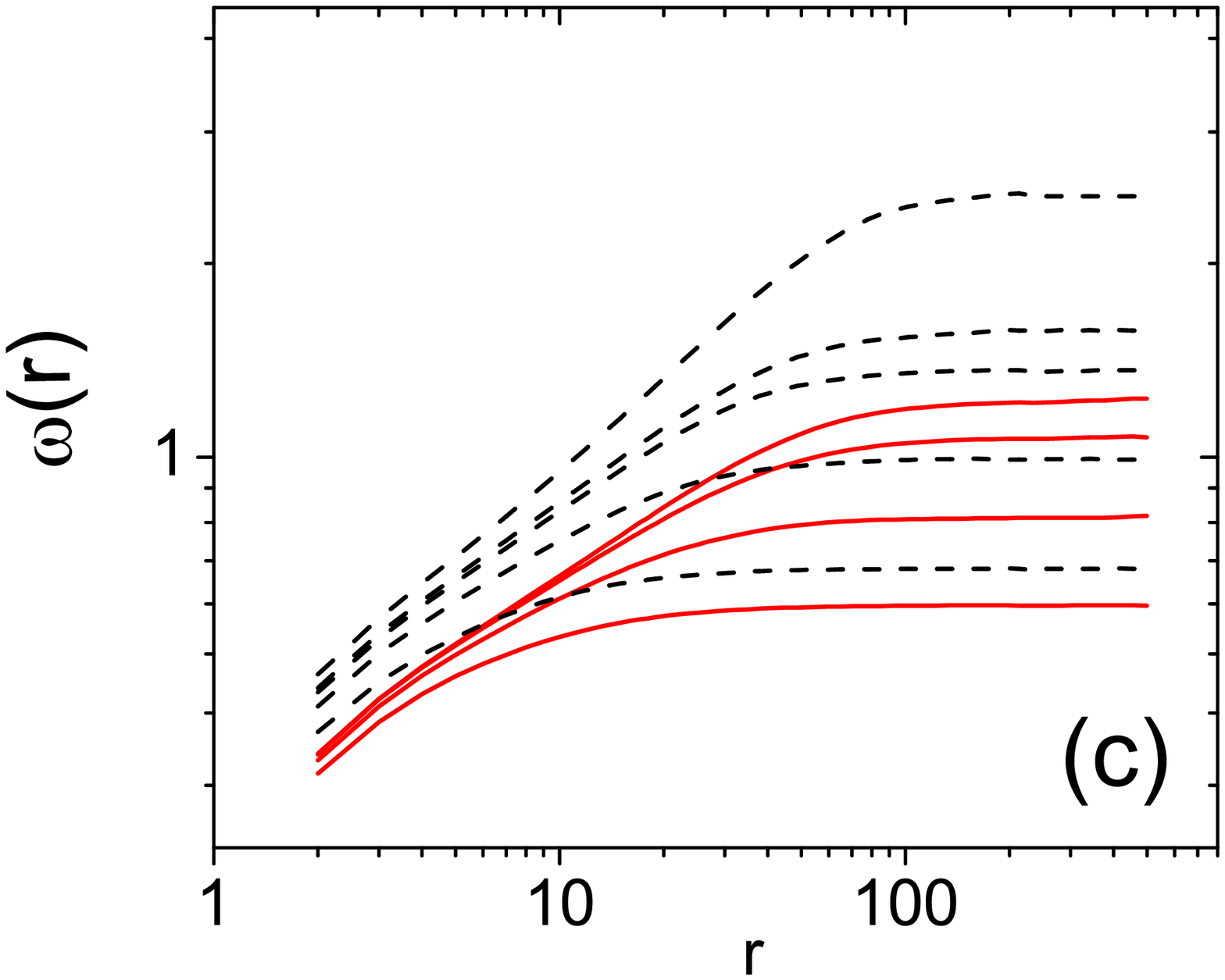}
\includegraphics [width=5.5cm] {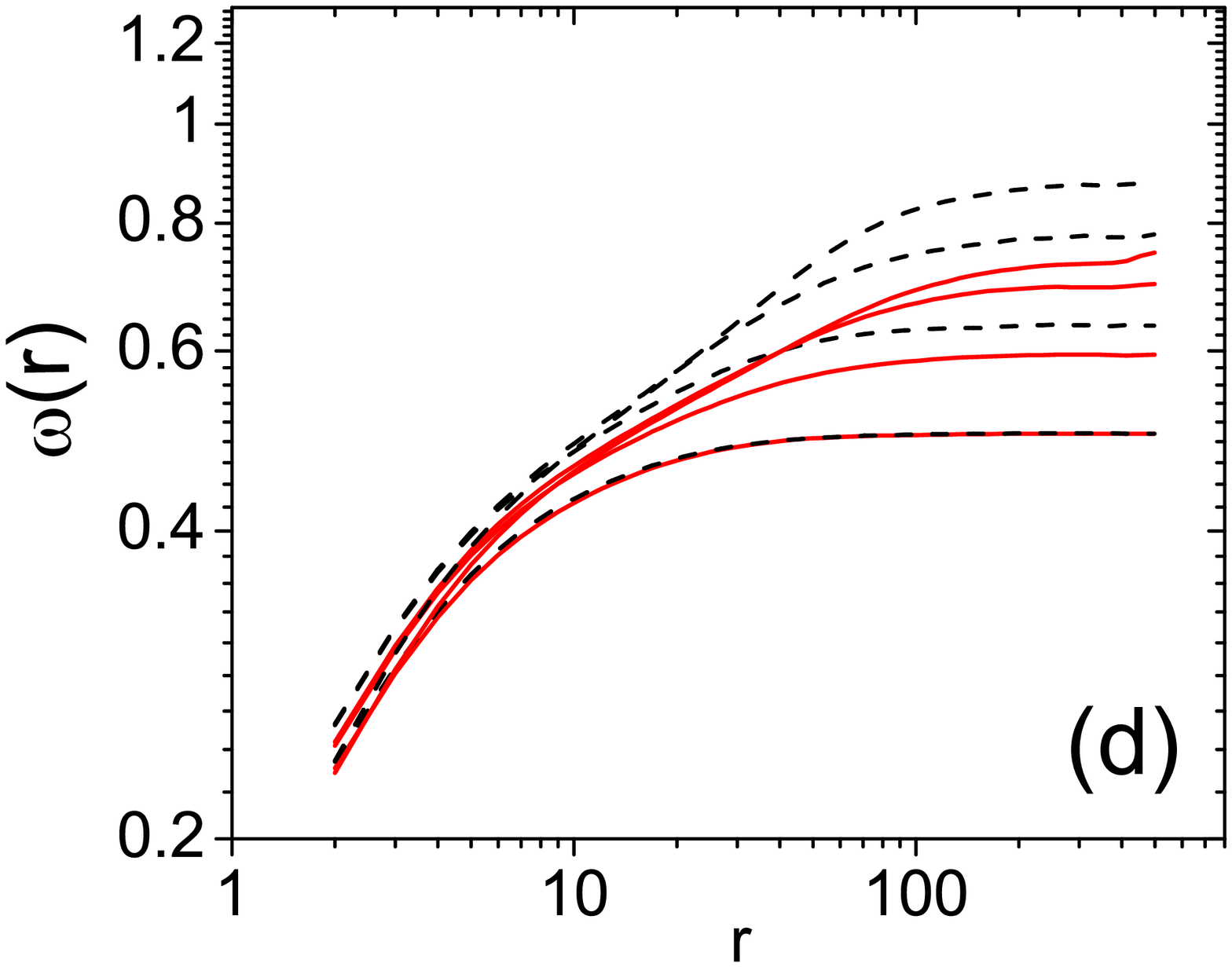}
\caption{Local roughness as a function of box size for (a) $R = 10$ , (b) $R=10^2$, (c) $R=10^{3}$,
and (d) $R=10^{4}$. Dashed (black) and full (red) lines indicate $\epsilon = 0.01$ and $0.1$,
respectively. For each set of parameters, from bottom to top, deposition times are $t=10$,
${10}^2$, $5\times {10}^2$, and ${10}^3$. For $R\leq10^{3}$ and $\epsilon = 0.01$, $t=10^{4}$ is also included .}
\label{wlocal}
\end{center}
\end{figure}
The effects of $R$ and $\epsilon$ on the roughness are very different from their effects
on island size in the submonolayer regime, in which scaling variables of the form
$R\epsilon^m$ appear in the crossovers (with rational $m>1$) \cite{fabiosub}.
Previous works on the basic CV model suggested deviations from VLDS scaling which
can be addressed in the light of our results.
The results of Figs. \ref{wglobal}a and \ref{wglobal}b are similar to those shown
in Ref. \protect\cite{kotrla1996}, which worked in the range ${10}^2< R<{10}^4$.
That work suggested the possibility of non-universal exponents due to the small slopes
of the fits of large $R$ data, but, as discussed above, roughness scaling is not expected
to appear in these conditions.
Temperature-dependent exponents $\beta$ were also suggested in Ref. \protect\cite{meng},
which simulated the model with $R\lesssim {10}^2$ up to $\approx {10}^3$
monolayers. A possible crossover from uncorrelated deposition \cite{barabasi} ($\beta =0.5$)
to Edwards-Wilkinson  scaling \cite{ew} ($\beta =0$) was suggested. However,
the former was obtained for $R\lesssim 1$, in which most atoms cannot execute a single step,
and the latter was obtained for very small values of the roughness, before the scaling regime.

\section{Local roughness scaling}
\label{local}

Figs. \ref{wlocal}a-d show the local roughness for some values of $R$ and $\epsilon$.
The results for small $R$ ($100$ or less) and small thicknesses ($1000$ monolayers
or less) show features of anomalous roughening, i. e. the $\log{w}\times\log{r}$
curves for different times are split for small $r$.
This was already observed for $\epsilon =0$ in Ref. \protect\cite{tempvar}.

For $R\geq {10}^3$, the curves for short times are also split, but coincide after
$\approx 100$ monolayers. The values of the local roughness are also very small,
thus they would be hard to be distinguished in an experiment.
Thus, the anomalous features disappear for
typical MBE temperatures ($R\geq {10}^5$); in this case, the presence of anomalous
scaling in experimental data is a clear indication of the presence of energy barriers
at step edges or other mechanisms that prevent surface smoothing by diffusion.

\begin{figure}
\begin{center}
\includegraphics [width=5.5cm] {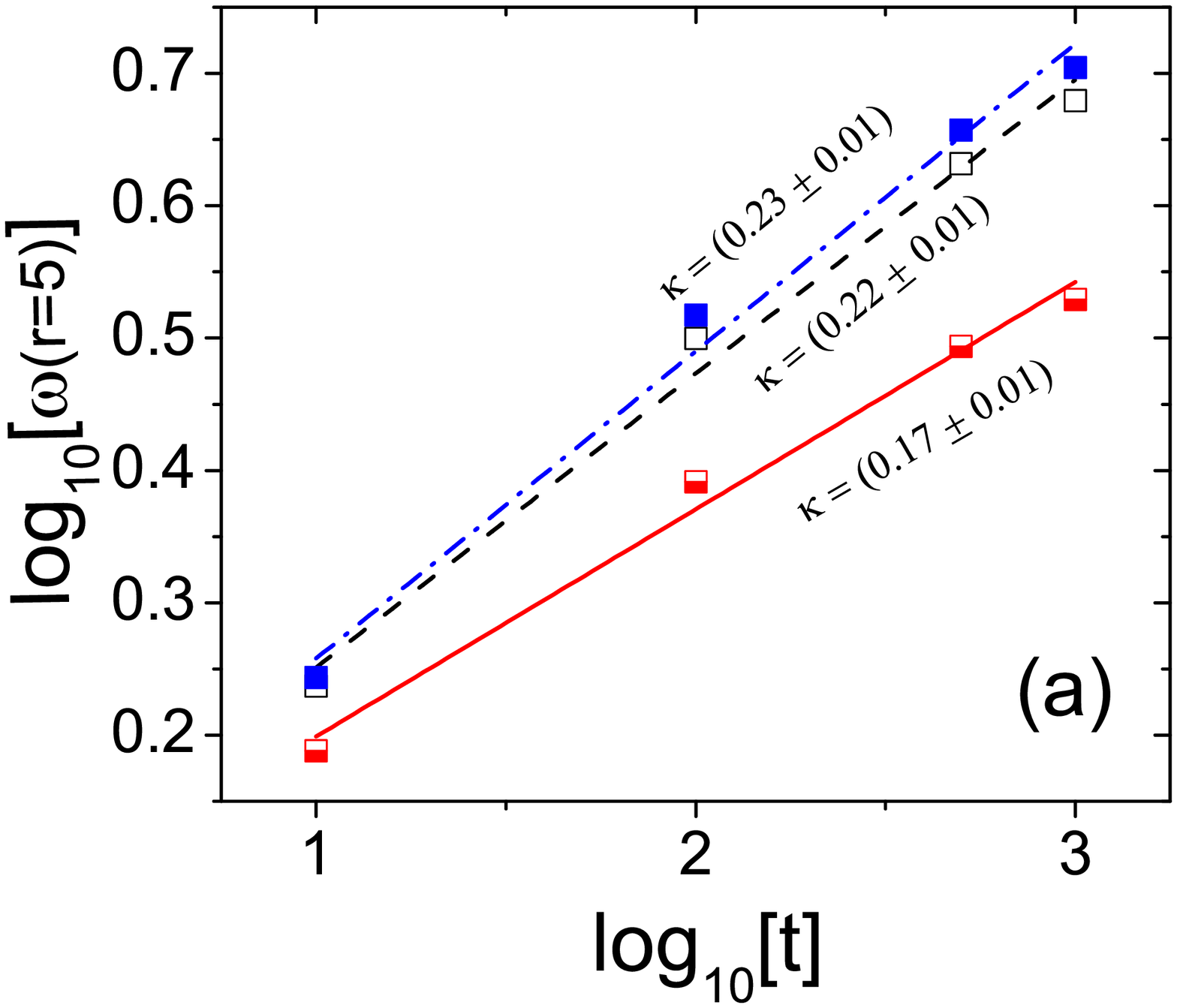}
\includegraphics [width=5.5cm] {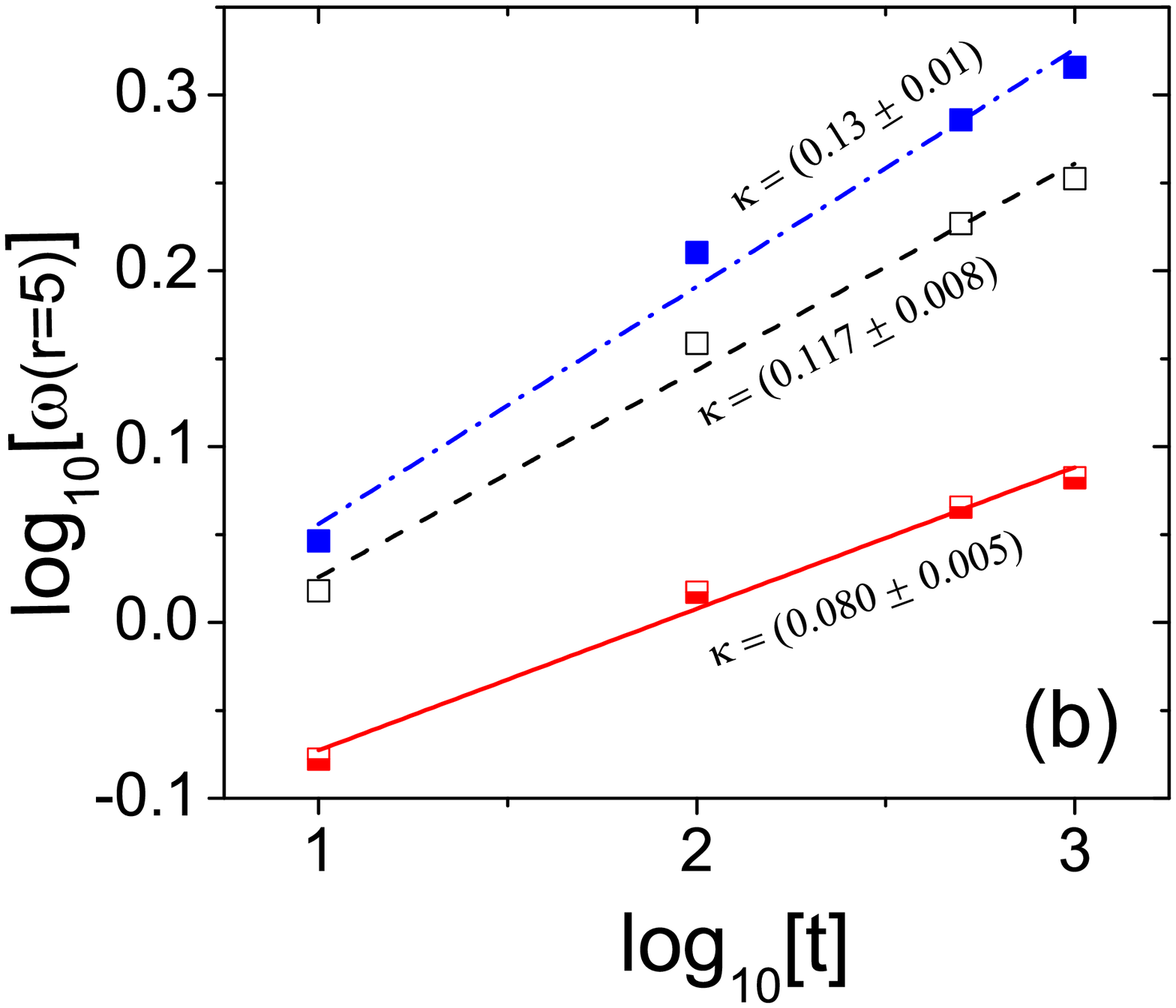}
\includegraphics [width=5.5cm] {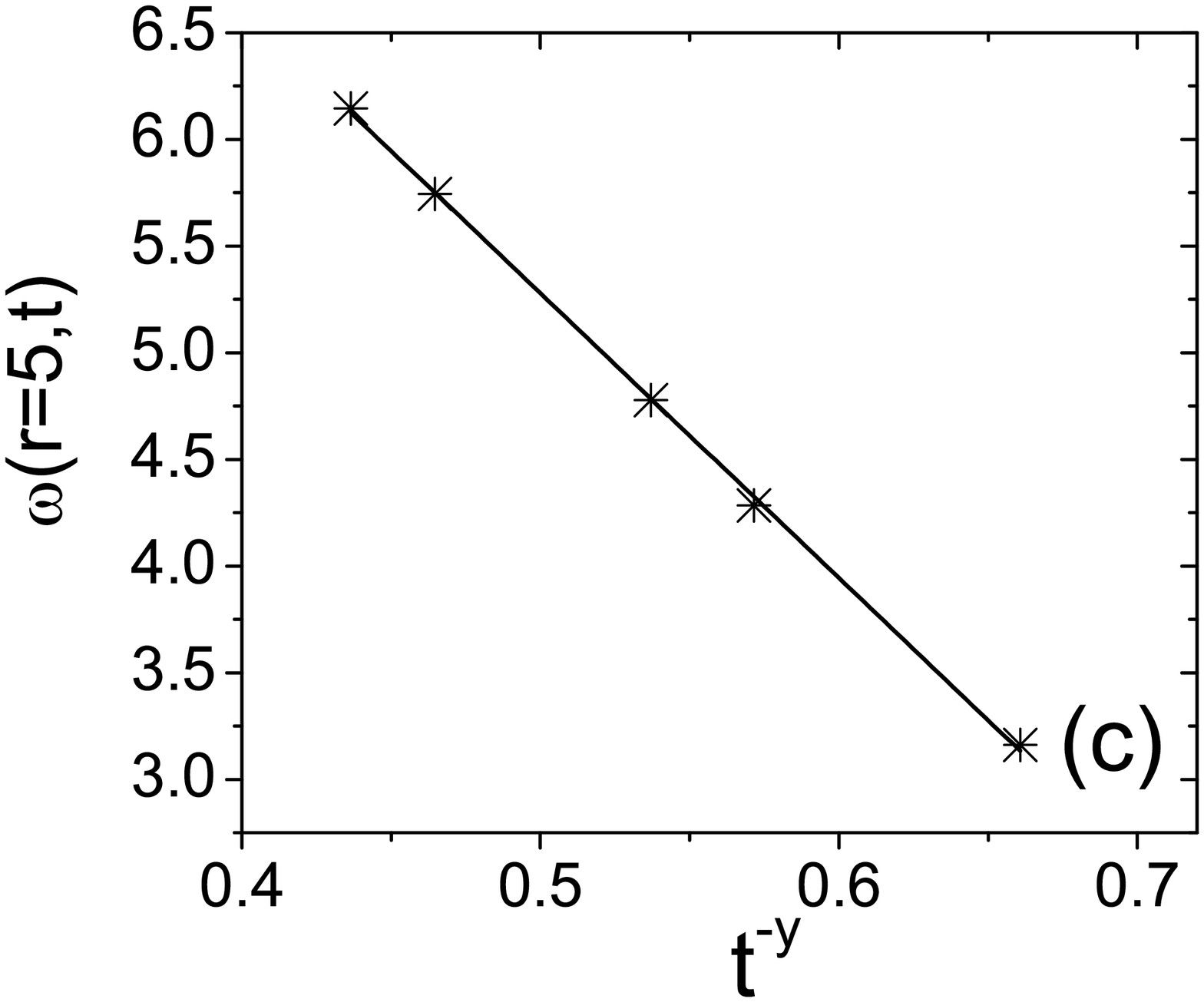}
\caption{(a), (b): time evolution of the local roughness for fixed box size, $r=5$, respectively
for $R=10$ and $R=10^2$; detachment parameters are $\epsilon =0$ (blue full squares),
$\epsilon =0.01$ (black open squares), and $\epsilon =0.1$ (red half squares); solid lines
are linear fits with the effective anomaly exponents shown in the plot.
(c) Local roughness at $r=5$ as a function of $t^{-y}$ for $R=10$
and $\epsilon =0.01$, using $y = 0.09$.}
\label{kappa}
\end{center}
\end{figure}

Figs. \ref{kappa} (a) and (b) show the time evolution of the roughness for fixed box
size, $r=5$, in growth with small $R$ and thicknesses between $10$ and $1000$.
Linear fits of those plots give effective anomaly exponents in the
range $0.08\leq \kappa\leq 0.23$. This procedure parallels the ones used in
experimental works. It confirms that thin films grown at very low
temperatures may show anomalous scaling features even in the absence of step edge
barriers.

However, this anomaly is only apparent, similarly to what occurs in other
models in the VLDS class \cite{anomvlds}. It is related to slowly vanishing
(instead of increasing) terms in the scaling of the local slopes or of the
small box local roughness, as
\begin{equation}
w\left( r_0,t\right)\sim A + Bt^{-y} .
\label{defy}
\end{equation}
In Fig. \ref{kappa} (c), we show $w\left( 5,t\right)$ versus $t^{-y}$ for $R=10$
and $\epsilon =0.01$, using $y = 0.09$. The good linear fit confirms the asymptotically
normal scaling, but the small value of $y$  gives large corrections even at long
times \cite{anomvlds}.

The local roughness scaling does not provide reliable estimates of exponent $\alpha$.
For small $R$, the scaling region of the $\log{w}\times\log{r}$ plot is very short
(Fig. \ref{wlocal}a) and the approximate slope is much smaller than $\alpha\approx 0.67$
\cite{crsosreis}. For large $R$, a longer scaling region appears, but the slope is
also very small. On the other hand, the dynamic exponent $z$ can be estimated by
the procedure proposed in Ref. \protect\cite{chamereis}.
The first step is to calculate a
characteristic length $r_c$ which is proportional to the correlation length at
a given time $t$. This is obtained by defining $r_c$ as
\begin{equation}
w\left(r_c,t\right) = kW\left( t\right) ,
\label{defrc}
\end{equation}
where $W\left( t\right)$ is the global width and $k$ is a
constant. From Eqs. (\ref{defbeta}) and (\ref{fvlocal}), it is expected that
\begin{equation}
r_c \sim t^{1/z} .
\label{scalingrc}
\end{equation}
Here, we consider $k=0.7$ for calculating $r_c$.

\begin{figure}[!h]
\begin{center}
\includegraphics [width=8.5cm] {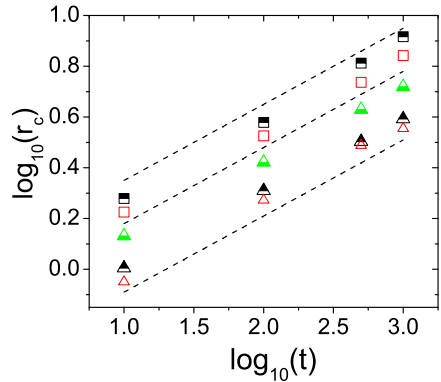}
\caption{Time evolution of $r_c$ for R=10 (triangles) and R=100 (squares). From bottom to top, detachment parameters are $\epsilon =0$, $0.01$, $0.1$ for R=10, and
$\epsilon = 0$, $0.01$ for R=100.
The dashed lines have slopes $0.3$.}
\label{rc}
\end{center}
\end{figure}
Fig. \ref{rc} shows the time evolution of $r_c$ for some values of $R$ and $\epsilon$;
they are restricted to $R\leq 100$ because the roughness was very small for larger $R$.
Linear fits give slopes between $0.30$ and $0.32$, which are in excellent agreement
with the VLDS value $1/z\approx 0.30$. It confirms that estimates of dynamic exponents
from the local roughness are more reliable than those of roughness exponents \cite{chamereis}.

\begin{figure}[!h]
\begin{center}
\includegraphics [width=8.5cm] {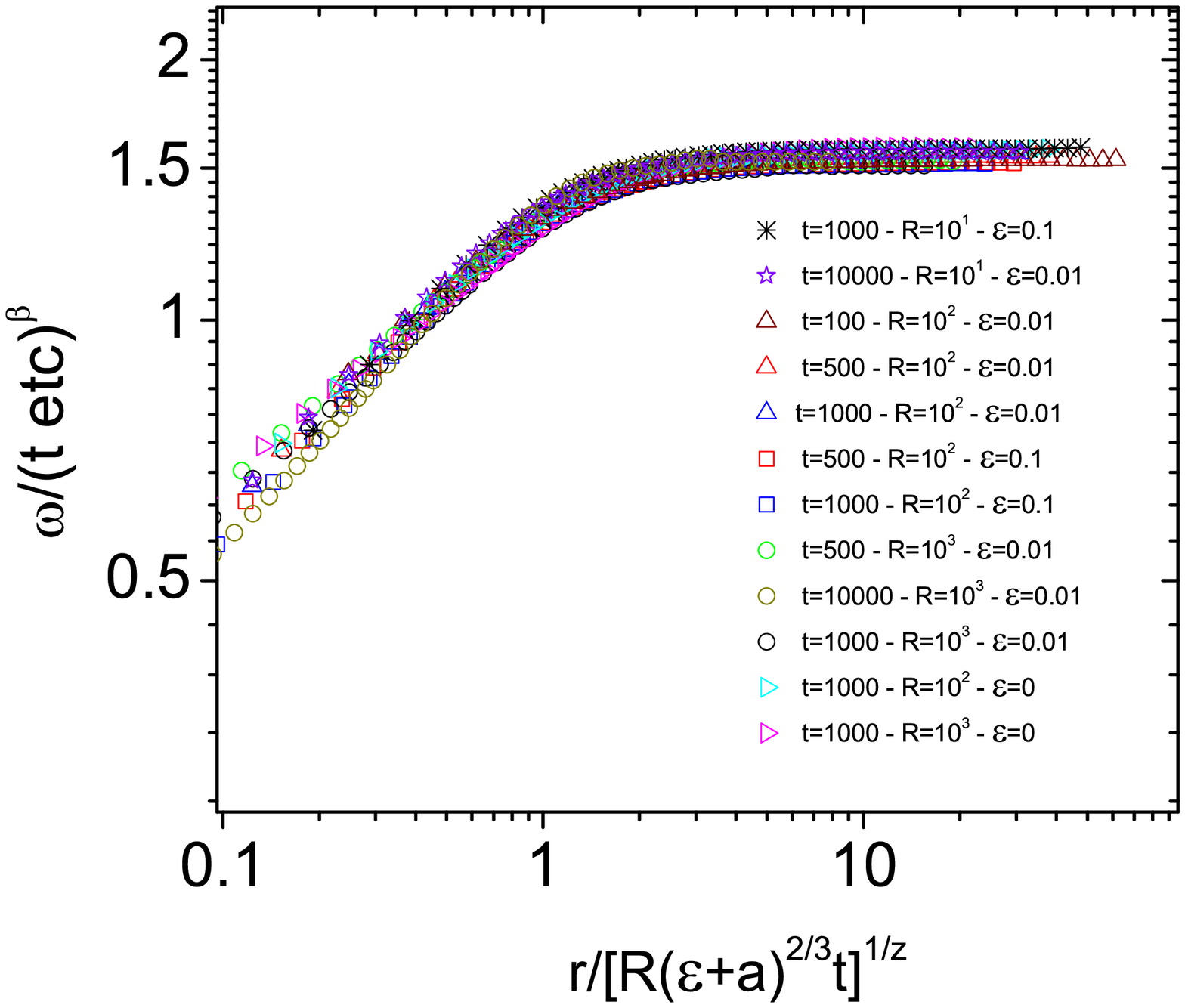}
\caption{Scaled local roughness as a function of scaled box size.}
\label{wscaling}
\end{center}
\end{figure}

The apparent anomaly for small $R$ does not affect the estimates of $z$ and $\beta$,
nor the global roughness scaling shown in Sec. \ref{global}.
Due to these weak corrections, we propose an scaling relation for
the local roughness considering: i) the general form of Family-Vicsek relation (\ref{fvlocal});
ii) the same temperature-dependent variable
$R^{3/2}\left( \epsilon + a\right)$ of the relation (\ref{WCV});
iii) the scaling of the correlation length consistent with $\xi\sim {\left( Rt\right)}^{1/z}$
for the case $\epsilon =0$ \cite{cdia}. This leads to
\begin{equation}
w\left( r,t\right) = \left[ \frac{t}{R^{3/2}\left( \epsilon +a\right)} \right]^\beta
g\left[ \frac{r}{{\left( R{\left( \epsilon +a\right)}^{2/3}t\right)}^{1/z}}\right],
\label{wCV}
\end{equation}
where $g$ is a scaling function.
Fig. \ref{wscaling} shows
$w/\left[ t / \left( R^{3/2}\left( \epsilon +a\right)\right) \right]^\beta$ as
a function of $r/{\left[ {R\left( \epsilon +a\right)}^{2/3}t\right]}^{1/z}$,
for several values of $R$, $\epsilon$, and $t$
(data with very small roughness is excluded, typically for short times or large $R$).
The data collapse is also good, showing that Eq. (\ref{wCV}) contains the leading
temperature-dependent terms of the local roughness scaling.

\section{Conclusion}
\label{conclusion}

We performed simulations of the basic CV model (without extra energy barriers at edges)
in $2+1$ dimensions for several values of the diffusion-to-deposition ratio $R$ and
detachment probability $\epsilon$ in order to study the scaling of global and local roughness.
Relatively short times consistent with thin film growth were considered.

The exponents $\beta$ ($z$) calculated from the global (local) roughness
confirm VLDS scaling with weak corrections.
The scaling variable $R^{3/2}\left( \epsilon + a\right)$, with $a=0.025$, represents
the temperature effects in the dynamic scaling relations, with excellent accuracy for
the global roughness (time scaling) and small corrections for the local roughness (time and
box size scaling). This shows that $R$ is the most important parameter to determine
the surface morphology, with much smaller effects of $\epsilon$.
This result for thin films is very different from that in submonolayer growth,
in which the scaling of island size and related quantities combines rational powers
of $R$ and $\epsilon$ \cite{fabiosub}. On the other hand, the present scaling variables involving
$R$ and $\epsilon$ are consistent with results of renormalization studies
\cite{haselPRE2008}.

The local roughness for $R\leq {10}^2$ shows evidence of anomalous scaling in the
range of thicknesses considered here. However, asymptotic normal scaling is observed,
with huge corrections expected even at long times. This is consistent with other VLDS
models.

Recent works on models with energy barriers for hopping across steps
determined exponents $\beta\approx 0.31$ and $1/z\approx 0.22$ \cite{lealJPCM,lealJSTAT}
for a range of temperatures; these values are quite different from the VLDS exponents.
The present model is of limited applicability to real solid films because they usually
show this type of additional energy barriers.
However, the methods proposed here may be useful for studying those extended models,
for instance incorporating the effect of other temperature-dependent
variables in the dynamic scaling, or investigating the question of
anomalous versus normal roughening.

\section*{Acknowledgements}

FDAAR acknowledges support from CNPq and FAPERJ (Brazilian agencies).
TAdA also acknowledges CNPq under grant 150874/2014-6.

\newpage


%

\end{document}